%% file: main_ieeeconf.tex
\documentclass[letterpaper, 10 pt, conference]{ieeeconf}
\IEEEoverridecommandlockouts
\usepackage[T1]{fontenc}
\usepackage{amsmath,amssymb,amsthm}
\usepackage{graphicx}
\usepackage{cite}
\usepackage{xcolor}
\usepackage{hyperref}

\newcommand{\Sres}{\text{Sres}}

\newcommand{\LC}{\text{LC}}
\newcommand{\Z}{\mathbb{Z}}
\newcommand{\R}{\mathbb{R}}
\newcommand{\Q}{\mathbb{Q}}
\newcommand{\C}{\mathbb{C}}
\newcommand{\blue}[1]{{\color{blue} #1}}

\newtheorem{theorem}{Theorem}

\newtheorem{proposition}[theorem]{Proposition}
\newtheorem{corollary}[theorem]{Corollary}
\theoremstyle{definition}
\newtheorem{definition}[theorem]{Definition}

\title{\LARGE \bf Reliable Computation of the $L_{\infty}$-Norm of Linear Time-Invariant Systems Depending on One Parameter}

\author{Florent Corniquel$^{1}$, 
Alban Quadrat$^{2}$, Fabrice Rouillier$^{2}$ and Grace Younes$^{3}$
\thanks{$^{1}$ Florent Corniquel is with Sorbonne Universit\'e and Universit\'e Paris Cit\'e, 
CNRS, IMJ-PRG, Inria Paris, France, {\tt florent.corniquel@inria.fr}} 
\thanks{$^{2}$ A. Quadrat and F. Rouillier are with PACE: 
Sorbonne Universit\'e and Universit\'e Paris Cit\'e, 
CNRS, IMJ-PRG, Inria Paris, France
{\tt \{alban.quadrat, fabrice.rouillier\}@inria.fr}}
\thanks{$^{3}$ G. Younes is with Sorbonne University Abu Dhabi, SAFIR, Abu Dhabi, UAE, {\tt grace.younes@sorbonne.ae}}
}

\begin{document}

\maketitle
\thispagestyle{empty}
\pagestyle{empty}

\begin{abstract}
The $L_{\infty}$-norm of a linear time-invariant system is a central quantity in robust control, used in robust stability tests, stability margins, and $H_{\infty}$ controller design. Physical models depend on parameters, such as masses, stiffnesses, or gains, so the $L_{\infty}$-norm becomes a function of these parameters. This computation reduces to studying the real critical points of a plane algebraic curve whose coefficients depend on the parameters. Previous approaches based on cylindrical algebraic decomposition describe the $L_{\infty}$-norm on the whole parameter space, but their cost grows very quickly with the number of parameters. In this paper, we propose instead to compute a generic parametrisation of the solutions, valid outside an explicitly characterised set of non-generic parameter values. In the case of one parameter, this parametrisation is obtained by interpolation from a certified solver for the non-parametric case. The resulting algorithm is fully certified, relying on a priori bounds and multimodular computations rather than on Gr\"obner bases. It is illustrated on examples from the robust control literature.
\end{abstract}

\section{Introduction}\label{sec:intro}

The $L_\infty$-norm of a linear time-invariant system, also known as its $H_\infty$-norm when the system is stable, measures the worst-case amplification of the energy of an input signal. Since the work of Zames \cite{zames1981feedback}, it has been at the heart of robust control theory: robust stability tests, stability margins, worst-case effects of disturbances, and $H_\infty$ controller design all rely on it \cite{Doyle,Zhou}. Unlike the $H_2$-norm, it has no closed-form expression in terms of the coefficients of the system. It is computed numerically, by testing whether a Hamiltonian matrix has eigenvalues on the imaginary axis and bisecting on the level $\gamma$ \cite{Boyd}, or by quadratically convergent variants of this scheme \cite{Boyd2,Bruinsma}. These algorithms are implemented in standard control software and perform well on most practical systems. Being based on floating-point arithmetic, however, they cannot certify their results, and some decisions, such as whether a pole lies on the imaginary axis, which makes the norm infinite, can only be made with respect to a tolerance.

Symbolic-numeric methods offer a certified alternative. Following Kanno and Smith \cite{kanno2006validated}, the computation of the $L_\infty$-norm can be reformulated as the search for the highest real critical point of a plane algebraic curve. This problem can be solved exactly, with certified numerical enclosures, by computer algebra methods: rational univariate representations, subresultants and certified real root isolation \cite{bouzidi2021computation,younes2025symbolic,JACA2026}. The most recent version of this approach builds the curve directly from state-space data, decides exactly whether a pole lies on the imaginary axis, and computes the norm at a cost comparable to that of standard numerical routines on most systems of the {\tt COMPleib} benchmark library \cite{JACA2026}.

The coefficients of a model usually depend on physical parameters, such as masses, stiffnesses, damping ratios or gains, which are uncertain or remain to be tuned, and the $L_\infty$-norm becomes a function of these parameters. Describing this function is a natural question in robust control, for instance, to understand how a stability margin degrades when a physical parameter varies, but few methods address it. The supremum of the real roots of a parametric univariate polynomial was studied in \cite{chen2013computing}, and parametric $H_\infty$ problems were considered in \cite{rance2018commande} for an application to gyro-stabilised sights. In \cite{QRY2024}, we described the norm as a function of the parameters using discriminant varieties and cylindrical algebraic decomposition. This gives a complete description of the parameter space, but at a cost that grows very quickly with the number of parameters.

In this paper, we follow a different route. Instead of decomposing the parameter space beforehand, we compute a \emph{generic parametrisation} of the solutions of the polynomial system that characterises the norm: a description of the solutions as functions of the parameters, valid outside an explicitly characterised set of non-generic parameter values. We will describe this object for an arbitrary number of parameters, and give an algorithm and its implementation in the case of one parameter. In this case, the generic parametrisation is obtained by interpolation from the certified solver of the non-parametric case, and the whole algorithm is certified, relying on a priori bounds and multimodular computations rather than on Gr\"obner bases. With one parameter, the norm, as a function of the parameter, describes a curve. The approach of \cite{QRY2024} describes this curve implicitly and projects it: each branch is identified by the index of a root of a univariate polynomial, and then by the position of the highest branch of the curve above it. With the generic parametrisation, the curve is described by a parametrisation together with a single point, which considerably reduces the size of the final description and the number of cases.

Section~\ref{sec:problem} recalls the non-parametric case and states the parametric problem, Section~\ref{sec:generic} introduces the generic parametrisation, Section~\ref{sec:algorithm} gives the algorithm for one parameter, and Section~\ref{sec:experiments} presents the experiments.

\section{Problem statement}\label{sec:problem}

This section recalls how the computation of the $L_\infty$-norm of a time-invariant linear ordinary differential system reduces to a problem on a real plane algebraic curve, and then extends this formulation to systems whose coefficients depend on parameters.

\subsection{The $L_\infty$-norm of a linear system}

A finite-dimensional linear time-invariant system with $m$ inputs and $p$ outputs is described in the frequency domain by its transfer matrix $G \in \R(s)^{p \times m}$, whose entries are real rational functions of the Laplace variable $s$. Let 
$\bar{\sigma}(M)$ denote the largest singular value of a complex matrix $M$.

Following \cite{Doyle,Zhou}, we denote by $RH_\infty$ the algebra of proper rational functions with real coefficients and no poles in 
$\overline{{\mathbb{C}_+}}:= \{s \in \C \mid \Re(s) \geq  0\}$, which are the transfer functions of stable systems, and by $RL_\infty$ the algebra of proper real rational functions with no poles on the imaginary axis $i \, \R$. For a matrix $G$ whose restriction to the imaginary axis has entries in $RL_\infty$, the $L_\infty$-norm is defined by
\begin{equation}\label{eq:Linf}
  \|G\|_\infty := \sup_{\omega \in \R} \bar{\sigma}\bigl(G(i \, \omega)\bigr).
\end{equation}
When the entries of $G$ belong to $RH_\infty$, the maximum modulus principle shows that \eqref{eq:Linf} coincides with $\sup_{s \in \mathbb{C}_+} \bar{\sigma}(G(s))$, the $H_\infty$-norm used in robust control \cite{Doyle,Zhou}. If $G$ has a pole on the imaginary axis, its $L_\infty$-norm is infinite.

\subsection{Algebraic characterisation}

Let $\widetilde{G}(s) := G^T(-s)$ denote the conjugate of $G$. On the imaginary axis, $\widetilde{G}(i \, \omega) = G(i \, \omega)^\ast$, so that the eigenvalues of $\widetilde{G}(i \, \omega)\,G(i \, \omega)$ are the squared singular values of $G(i \, \omega)$. This leads to the following characterisation.

\begin{proposition}[\cite{kanno2006validated}]\label{prop:Kanno-Smith}
Let $G \in \R(s)^{p \times m}$ be such that $G_{|i\R} \in RL_\infty^{p \times m}$, $\gamma > 0$, and $\Phi_\gamma(s) := \gamma^2 I_m - \widetilde{G}(s)\,G(s)$. Then, we have $\gamma > \|G\|_\infty$ if and only if $\gamma > \bar{\sigma}\bigl(G(i \, \infty)\bigr)$ and $\det \Phi_\gamma(i \, \omega) \neq 0$ for all $\omega \in \R$.
\end{proposition}

The determinant $\det \Phi_\gamma(i\omega)$ is a real rational function of $\omega$ and $\gamma$, and even a rational function of $\omega^2$ and $\gamma^2$. We write
\begin{equation}\label{eq:n_over_d}
  \det \Phi_\gamma(i \, \omega) = \frac{n(\omega,\gamma)}{d(\omega)},
\end{equation}
with $n \in \R[\omega,\gamma]$, $d \in \R[\omega]$ and $\gcd(n,d) = 1$.

\begin{definition}\label{def:curve_n}
The polynomial $n \in \R[\omega,\gamma]$ in \eqref{eq:n_over_d} is called the \emph{polynomial associated with the $L_\infty$-norm} of $G$.
\end{definition}

Since $G$ has no pole on the imaginary axis, $d$ has no real root, and $\det \Phi_\gamma(i \, \omega)$ vanishes exactly when $n(\omega,\gamma)$ does. By Proposition~\ref{prop:Kanno-Smith}, $\|G\|_\infty$ is thus the largest $\gamma$ for which the equation $n(\omega,\gamma) = 0$ has a real solution $\omega$, possibly in the limit $\omega \to \infty$. Geometrically, every real point of the curve
\[
  \mathcal{C} := \{(\omega,\gamma) \in \R^2 \mid n(\omega,\gamma) = 0\}
\]
has a $\gamma$-coordinate bounded by $\|G\|_\infty$, and the norm is reached either at a critical point of $\mathcal{C}$ or along a horizontal asymptote. This is made precise below, where $\pi_\gamma : \R^2 \to \R$ denotes the projection onto the $\gamma$-axis and $V_\R(\cdot)$ the set of real solutions of a polynomial system.

\begin{proposition}[\cite{kanno2006validated,bouzidi2021computation}]\label{prop:caracterization_infty_norm}
Let $G$ be as in Proposition~\ref{prop:Kanno-Smith}, $n$ the polynomial associated with its $L_\infty$-norm, and $\bar{n}$ the square-free part of $n$. Then
\[
  \|G\|_\infty = \max \Bigl( \pi_\gamma\Bigl(V_\R\bigl(\bar{n}, \tfrac{\partial \bar{n}}{\partial \omega}\bigr)\Bigr) \ \cup \ V_\R\bigl(\LC_\omega(\bar{n})\bigr) \Bigr),
\]
where $\LC_\omega(\bar{n}) \in \R[\gamma]$ is the leading coefficient of $\bar{n}$ with respect to $\omega$.
\end{proposition}

The computation of $\|G\|_\infty$ is thus reduced to two tasks: finding the largest real root of the univariate polynomial $\LC_\omega(\bar{n})$, which accounts for the horizontal asymptotes of $\mathcal{C}$, and finding the largest $\gamma$-coordinate among the real solutions of the bivariate polynomial system
\begin{equation}\label{eq:Sigma}
  \Sigma := \Bigl\{\, \bar{n}(\omega,\gamma) = 0, \ \frac{\partial \bar{n}}{\partial \omega}(\omega,\gamma) = 0 \,\Bigr\}.
\end{equation}

\subsection{Certified computation in the non-parametric case}\label{subsec:nonparam}

When $n$ has rational coefficients, the system $\Sigma$ in \eqref{eq:Sigma} can be solved exactly. Several approaches have been developed and compared in \cite{bouzidi2021computation}. The first ones put $\Sigma$ in generic position by a linear change of coordinates $t = \gamma + a \, \omega$, chosen so that no two solutions have the same projection on the new axis \cite{bouzidi2015separating,bouzidi2016solving}. A Rational Univariate Representation (RUR) of the system in the new coordinates then gives isolating boxes for all its real solutions, from which the largest $\gamma$-coordinate is read. A variant uses a separating form that preserves the order of the projections \cite{cheng2009root}, at the price of larger coefficients. A second approach isolates the real roots of the resultant $\operatorname{Res}(n, \partial n/\partial\omega, \omega) \in \Q[\gamma]$ and certifies, by counting real roots with Sturm--Habicht sequences \cite{gonzalez1998sturm}, which of them is the $\gamma$-coordinate of a real solution of $\Sigma$. In the {\tt Maple} implementations of \cite{bouzidi2021computation}, this real-root counting approach was the most efficient, but recent progress in the computation of RURs \cite{demin2024reading} has changed this comparison.

The most recent pipeline \cite{JACA2026} builds on these tools and reduces the size of the system handed to the solver in four ways.
\begin{enumerate}
  \item \emph{From the state-space.} When $G = C(sI - A)^{-1}B + D$ is given by a realization, the curve is built directly from $(A, B, C, D)$ by a Faddeev--Leverrier recurrence, without forming the transfer matrix $G$, and from the smaller of the two Gram matrices $\widetilde{G} \, G$ and $G \, \widetilde{G}$, so that its degree in $\gamma^2$ is $\min(m,p)$.
  \item \emph{Exact decision on the imaginary axis.} Modes of the realization that cancel in $G$ are removed by a gcd, and the presence of poles on the imaginary axis, for which $\|G\|_\infty = \infty$, is decided by a second gcd and a real-root count, without any tolerance.
  \item \emph{Halving the degrees.} Since $n$ only depends on $\omega^2$ and $\gamma^2$, we write $n(\omega,\gamma) = c(\Omega,\Gamma)$ with $\Omega = \omega^2$ and $\Gamma = \gamma^2$. As $\partial n/\partial\omega = 2\omega\, (\partial c/\partial \Omega)(\omega^2,\gamma^2)$, the system $\Sigma$ becomes, once the factor $\omega$ is removed,
  \begin{equation}\label{eq:Sigma_reduced}
    c(\Omega,\Gamma) = 0, \qquad \frac{\partial c}{\partial \Omega}(\Omega,\Gamma) = 0,
  \end{equation}
  which halves the degree in each variable; it is solved after dividing $c$ and $\partial c/\partial\Omega$ by their gcd. Removing the factor $\omega$ discards the critical points at $\omega = 0$, and $\Omega=0$ becomes the boundary of the domain $\Omega \ge 0$. The candidates for $\|G\|_\infty^2$ are therefore the $\Gamma$-coordinates of the real solutions of \eqref{eq:Sigma_reduced} with $\Omega \ge 0$, the real roots of $c(0,\Gamma)$ for $\omega = 0$, and the real roots of $\LC_\Omega(c)$ for $\omega \to \infty$; $\|G\|_\infty^2$ is the largest of them \cite{JACA2026}.
  \item \emph{Eliminating $\Gamma$.} The system \eqref{eq:Sigma_reduced} is solved by computing the subresultant sequence of the two reduced polynomials with respect to $\Gamma$, the degree of $c$ in $\Gamma$ being only $\min(m,p)$, and the triangular decomposition it induces \cite{RLP2017}. A rational parametrisation of the solutions is read off this decomposition, and the real roots of the eliminant in $\Omega$ are isolated, which gives a certified enclosure of the norm together with the frequency where it is attained.
\end{enumerate}

In the rest of the paper, we use the notation of this pipeline: $x:= \Omega$ and $y:= \Gamma$, and $P$, $Q$ denote $c$ and $\partial c/\partial \Omega$ divided by their gcd, with integer coefficients after clearing denominators. They are coprime, and the non-parametric solver amounts to solving $\{P = 0, Q = 0\}$ by eliminating $y$ through the subresultant sequence of $P$ and $Q$. The boundary and asymptotic candidates are read off $c$ itself.
We now consider the case where the coefficients of $G$, hence those of $n$, depend on parameters.

\subsection{The parametric case}\label{subsec:param}

We now consider transfer matrices whose entries depend rationally on real parameters $\mathbf{w} := (w_1, \dots, w_s)$. The polynomial associated with the $L_\infty$-norm then belongs to $\Q[\mathbf{w}][\omega,\gamma]$ (we assume it is square-free, replacing it by its square-free part if needed), and the norm becomes a function of $\mathbf{w}$. Let
\[
\begin{array}{l}
  \Sigma := \Bigl\{ (\omega,\gamma,\mathbf{w}) \in \R^{2+s} \ \Big| \ n = 0, \ \dfrac{\partial n}{\partial \omega} = 0 \Bigr\}, \\[2ex]
  \Sigma_\infty := \bigl\{ (\gamma,\mathbf{w}) \in \R^{1+s} \ \big| \ \LC_\omega(n) = 0 \bigr\}.
\end{array}
\]
By Proposition~\ref{prop:caracterization_infty_norm}, for generic values of $\mathbf{w}$ at which $G$ has no pole on the imaginary axis, $\|G\|_\infty$ is the largest $\gamma$-coordinate of the points of $\Sigma$ and $\Sigma_\infty$ above $\mathbf{w}$.

\begin{center}
\fbox{\parbox{\dimexpr\columnwidth-2\fboxsep-2\fboxrule\relax}{\textbf{Problem statement.} Let $G$ be a transfer matrix depending on real parameters $\mathbf{w} = (w_1, \dots, w_s)$, and let $n \in \Q[\mathbf{w}][\omega,\gamma]$ be the polynomial associated with its $L_\infty$-norm. Describe, as a function of $\mathbf{w}$,
\[
  \|G\|_\infty = \max \Bigl( \pi_\gamma\Bigl(V_{\R}\bigl(n, \tfrac{\partial n}{\partial \omega}\bigr)\Bigr) \ \cup \ V_\R\bigl(\LC_\omega(n)\bigr) \Bigr),
\]
that is, the largest $\gamma$-coordinate of the real solutions of $\Sigma$ and $\Sigma_\infty$ above each value of the parameters.}}
\end{center}

In the variables $x = \Omega$ and $y = \Gamma$ of Section~\ref{subsec:nonparam}, $\|G\|_\infty^2$ is thus, for generic values of $\mathbf{w}$, the largest of the following candidates: the $y$-coordinates of the real solutions with $x \ge 0$ of the system $\{P = 0, Q = 0\}$, where $P, Q \in \Z[\mathbf{w}][x,y]$ are coprime; the real roots of $c(0,y)$, for $\omega = 0$; and the real roots of $\LC_x(c)$, for $\omega \to \infty$. The last two families are univariate in $y$ for each value of $\mathbf{w}$. The parametric difficulty therefore lies in the bivariate system $\{P = 0, Q = 0\}$, which is the object of Section~\ref{sec:generic}.

\paragraph{Parametric systems.} Questions of this kind, where one wants to describe how the solutions of a polynomial system depend on its parameters, are classically addressed with the notion of \emph{discriminant variety} \cite{lazard2007solving}. For a system that has finitely many solutions, all simple, for almost all parameter values, a discriminant variety is a set of ``bad'' parameter values, of dimension less than $s$, outside of which the solutions behave regularly: no solution goes to infinity, no two solutions meet, and the number of real solutions is constant on each connected component of its complement. These components, or \emph{cells}, can be described by a cylindrical algebraic decomposition (CAD) \cite{collins1975quantifier}. Solving the system at one sample point of each cell then gives the behaviour of the solutions on the whole cell.

\paragraph{Previous approach.} This strategy was applied to the $L_\infty$-norm in \cite{QRY2024}. The main steps are:
\begin{enumerate}
  \item compute the resultant $R := \operatorname{Res}(n, \partial n/\partial \omega, \omega) \in \Q[\mathbf{w}][\gamma]$, whose real roots contain the candidate values of the norm;
  \item compute a discriminant variety of $\{R = 0\}$ with respect to the parameters, essentially given by $\operatorname{Res}(R, \partial R/\partial \gamma, \gamma) \in \Q[\mathbf{w}]$ together with the constraints on the parameters: outside of it, the real roots of $R$ are continuous functions of $\mathbf{w}$ whose graphs do not cross;
  \item decompose its complement into cells with a CAD, and pick a sample point in each cell;
  \item at each sample point, solve the non-parametric problem and record the index of the norm in the sorted list of the real roots of $R$.
\end{enumerate}
Since the roots of $R$ keep the same order on each cell, the index recorded at the sample point gives the norm on the whole cell. For instance, for $G = (a \, s + 1)/(b \, s + 1)$ with $a, b > 0$ \cite{Doyle}, two cells are obtained, $0 < a < b$ and $a > b > 0$, on which the norm is respectively $1$ and $a/b$ \cite{QRY2024}.

This approach describes the norm on the whole parameter space, but its cost is that of a CAD, doubly exponential in the number of parameters. In practice, the number of cells grows quickly: a single mechanical mode with four parameters already yields $52$ cells, and a related system of higher degree yields $920$ \cite{QRY2024}.

\paragraph{Approach of this paper.} Rather than decomposing the parameter space, we compute a generic parametrisation of the solutions of $\{P = 0, Q = 0\}$, valid outside an explicit finite union of algebraic sets, the non-generic parameter values being characterised exactly (Section~\ref{sec:generic}). For one parameter, it is interpolated from the non-parametric solver of Section~\ref{subsec:nonparam}, used as a black box, and certified by a priori bounds and multimodular computations rather than by Gr\"obner bases, whose certification at a reasonable cost is not known.

\section{Computing a generic parametrisation}\label{sec:generic}

Consider $P,Q \in \Z[\mathbf{w}][x,y]$ two coprime polynomials depending on two variables and on parameters $\mathbf{w} := (w_1, \dots, w_s)$. We denote by $d_y$ the degree of $P$ in $y$. We are interested in solving the system $\{P = 0,  Q = 0 \}$
by expressing, when possible, the unknowns $x,y$ in terms of $\mathbf{w} \in \Q^s$.

\subsection{Subresultants sequence}\label{subsec:subres}

In the parametric case, we transpose the subresultants-based approach presented in \cite{RLP2017}. The idea is to compute the subresultants sequence $\Sres_y (P,Q) := (S_{i})_{i \in [\![0, d_y]\!]}$ between $P$ and $Q$ seen as elements of $\Z[\mathbf{w},x][y]$, where $[\![0, d_y]\!]$ is the set of integers between $0$ and $d_y$. We suppose that $\deg_y(P)\ge\deg_y(Q)$ and, when they are equal, we set $S_{d_y}:=P$. For each $i \in [\![0, d_y]\!]$, we write $S_i = \sum_{j=0}^{i} s_{i,j} (x ; \mathbf{w}) \, y^j$
where every $s_{i,j}$ is an element of $\Z[\mathbf{w},x]$. 

It is well known that the subresultant polynomial $S_i$ does not identically vanish if and only if its leading coefficient (that is $a_{i,i}$) is not identically zero. In that case, it has degree in $y$ exactly $i$. In particular, $R_y := S_0 \in \Z[\mathbf{w}][x]$ is the resultant of $P$ and $Q$ with regard to the variable $y$ and we denote it by $R_y$. It is non identically zero since $P$ and $Q$ are supposed to be coprime.

\subsection{Triangular decomposition}\label{subsec:triangular}
Denote by $i_0$ the index of the first non-identically zero subresultant polynomial (other than the resultant). Denote by $x_0$ a theoretical root of $R_y$ which naturally depends on $\mathbf{w}$. If $s_{i_0, i_0}$ does not vanish at $x_0$ then $S_{i_0}$ is the gcd of $P(x_0,y ; \mathbf{w})$ and $Q(x_0,y; \mathbf{w})$ (up to a factor in $\Z[w]$). Therefore, the roots of $S_{i_0} (x_0,y ; \mathbf{w})$ are the $y$-coordinates of the roots (they also naturally depend on $\mathbf{w}$). Thus, if $R_y$ and $s_{i_0,i_0}$ do not have any common root, we have the equivalence
\begin{equation}\label{equiv system}
\begin{cases} P(x,y ; \mathbf{w}) = 0, \\ Q(x,y ; \mathbf{w}) = 0,\end{cases} \iff \begin{cases} R_y (x; \mathbf{w}) = 0, \\ S_{i_0} (x,y ; \mathbf{w}) = 0.\end{cases}
\end{equation}

\subsection{Non generic parameters values}\label{subsec:non generic}
The natural question is now to characterise the values $w_0 \in \C^s$ that we can substitute in Equivalence \ref{equiv system} while keeping the equivalence true. The parameter values for which this is not the case are called \ \textit {non-generic}. We exclude the values $w_0 \in \C^s$ such that :

\begin{enumerate}
    \item $\operatorname{Res}(\LC_y(P),\LC_y(Q),x)(w_0)=0$. Those are parameters values for which $(S_{i} (x,y; w_0))_{i \in [\![0,d_y]\!]}$ is not the subresultants sequence of $P(x,y ; w_0)$ and $Q(x,y ; w_0)$;
    \item $\text{cont}_{x}(s_{i_0,i_0})(w_0) = 0$, where $\text{cont}_{x}$ is the content, that is, the $gcd$ of all the coefficients in $x$. Those are parameters values for which $s_{i_0, i_0}$ (and thus $S_{i_0}$) vanishes. In this case, $S_{i_0}$ is identically zero and the correct subresultant polynomial to consider would be one of higher degree in $x$.  
    \item $\operatorname{Res}(R_y,s_{i_0,i_0}, x) (w_0) = 0$. Those are the parameters values for which $R_y(x ; w_0)$ and $s_{i_0, i_0}$ have common roots in $x$. In this case, $S_{i_0}$ vanishes at some $x$-roots of $R_y(x ; w_0)$ making us miss some solutions.
\end{enumerate}

We denote by respectively $V_1, V_2$ and $V_3$ the sets of values $w_0 \in \C^s$ described just before. We summarise what has been said in the following statement.

\begin{theorem}\label{thm:generic}
	For all $w_0 \in \C^s \backslash (V_1 \cup V_2 \cup V_3)$, 
	\begin{equation*}
\begin{cases} P(x,y ; w_0) = 0, \\ Q(x,y ; w_0) = 0,\end{cases} \iff \begin{cases} R_y (x; w_0) = 0, \\ S_{i_0} (x,y ; w_0) = 0.\end{cases}.
\end{equation*}
\end{theorem}

\paragraph{Remarks} The convention for $S_{d_y}$ matters, even for scalar systems of order two. For $G=1/(s^2+2 \, \xi \, s+1)$, we have $P=y\,D(x)-1$ with $D(x)=(1-x)^2+4 \, \xi^2 \, x$, and $Q=y\,D'(x)$. Then $R_y=\pm D'(x)$ and, at its root $x_0=1-2 \, \xi^2$, $Q(x_0,y)$ is identically zero. With $S_{1}=P$, $s_{1,1}=D$ and $V_3\neq\C^s$; with $S_{1}=Q$, $s_{1,1}=D'$ vanishes at every root of $R_y$, and the solution $(x_0,1/D(x_0))$, which gives the norm, would be missed.

Theorem~\ref{thm:generic} is meaningful when $V_3\neq\C^s$, that is when, for generic $\mathbf{w}$, the gcd of $P(x_0,y)$ and $Q(x_0,y)$ has the same degree $i_0$ above all the roots $x_0$ of $R_y$. Otherwise, the roots of $R_y$ are split according to this degree $k$, by gcds of $\overline{R_y}$ with $s_{k,k}$, and the statement applies to each part with $S_{k}$; the algorithm of Section~\ref{sec:algorithm} reads these parts on the triangular decomposition.

\subsection{Rational parametrisation}\label{subsec:ratpara}

We can go one step further in the solving of the system by constructing a rational parametrisation. For that purpose, observe that if $S_{i_0}$ has degree one in $y$ (that is $i_0=1$), it is written $s_{1,1}(x ; \mathbf{w}) y + s_{1,0} (x; \mathbf{w})$ and as long as $s_{1, 1}$ and $R_y$ are coprime, Equivalence \eqref{equiv system} can be rephrased as 

\begin{equation}\label{rat param}
    \begin{cases}
        P(x,y ;\mathbf{w}) = 0, \\ Q(x,y ; \mathbf{w}) = 0,
    \end{cases} \iff \begin{cases}
        R_y(x ; \mathbf{w}) = 0, \\ y = -\frac{s_{1,0}(x ; \mathbf{w})}{s_{1,1}(x ; \mathbf{w})}.
    \end{cases}
\end{equation}

However, $S_{i_0}$ might not have degree one. We can still derive a similar parametrisation following the philosophy of \cite{demin2024reading}. 

\begin{definition}
    A linear form $t=a \, x+b \, y \in \Q[x,y]$ is \textit{generically separating} if there exists an algebraic set $V \subset \C^s$ such that for all $w_0 \in \C^s \backslash V$, $t$ takes different values on every distinct solution of $\{P(x,y ; w_0) = Q(x,y ; w_0)= 0\}$.
\end{definition}

We suppose that $x$ is generically separating and we denote by $V_4 \subset \C^s$ such an algebraic set. In this case, $V_4$ is the set of roots of $\operatorname{Disc}(\overline{R_y}) \subset \C^s$, the discriminant of the square-free part of $R_y$. For any $w_0 \in \C^s \backslash V_4$, to any root $x_0 \in \C$ of $R_y(x ; w_0)$ is associated a unique $y_0 \in \C$ such that $(x_0, y_0)$ is a root of the system. Therefore, $y_0$ is the only root of $S_{i_0}(x,y;w_0)$ and $S_{i_0} = s_{i_0,i_0} (y-y_0)^{i_0}$.

By comparing the coefficient of the binomial expansion with the coefficients of degree $i_0-1$ in the dense representation of $S_{i_0}$, we get that $-i_0 \, s_{i_0,i_0} \, y_0 = s_{i_0,i_0-1}$. That explains the following statement.

\begin{corollary}\label{coro:rur}
    For all $w_0 \in \C^s \backslash (V_1 \cup V_2 \cup V_3 \cup V_4)$, 
    \begin{equation*}
\begin{cases} P(x,y ; w_0) = 0, \\ Q(x,y ; w_0) = 0,\end{cases} \iff \begin{cases} R_y (x; w_0) = 0, \\ y = -\frac{s_{i_0,i_0-1} (x ; w_0)}{i_0 s_{i_0, i_0}(x ; w_0)}.\end{cases}
\end{equation*}
\end{corollary}

\subsection{Linear change of variables}\label{subsec:lin change}
If neither $x$ nor $y$ is generically separating, one can still obtain a similar parametrisation by substituting $x = t - a \, y$ in the original system if $t = x + a \, y$ is known to be generically separating, and then computing the subresultants sequence with respect to variable $y$. Variable $t$ then plays the role of $x$, and $y$ remains the eliminated variable.
For instance, in Section \ref{sec:algorithm}, such a linear form is obtained through a call to \cite[Algorithm 6]{demin2024reading}.

\section{Algorithm}\label{sec:algorithm}
\input{sec_algorithm}

\section{Experiments}\label{sec:experiments}
\input{sec_experiments}

\end{document}

%% file: sec_algorithm.tex
We now restrict to one parameter $w$, so that $P,Q\in\Z[w][x,y]$ with $x:=\Omega$ and
$y:=\Gamma$. The algorithm computes a generic parametrisation of the solutions of
$\{P=0,Q=0\}$ by interpolation in $w$ and multimodular lifting, using as a black box the
bivariate triangular decomposition~\cite{RLP2017} modulo a prime of the non-parametric solver of
Section~\ref{subsec:nonparam}. It then derives from this parametrisation a description of
$\|G\|_\infty$ on the stability domain of $G$.

\subsection{The object computed}

Let $a\in\Z$ and $t:=x+a \, y$, a linear form meant to separate the solutions. After the change of
variables $x\mapsto t-a \, y$, we still denote by $P$ and
$Q$ the polynomials $P(t-a \, y,y;w)$ and $Q(t-a \, y,y;w)$, and eliminate $y$. For a value $t_0$ of
$t$, the \emph{level} of $t_0$ is the degree in $y$ of $\gcd(P(t_0,y),Q(t_0,y))$. Let
$S_k=\sum_{j\le k}s_{k,j}(T;w)\,y^j$ be the subresultant of index $k$ of $P$ and $Q$ with respect
to $y$. Given a root $t_0$ of level $k$, the discussion held in Subsection \ref{subsec:ratpara} applies and
\[
  y_0=-\frac{s_{k,k-1}(t_0 ; w)}{k\,s_{k,k}(t_0 ; w)}.
\]
The output has the following form: for each level $k$ which is not empty for generic $w$, four polynomials of
$\Z[w][t]$ written
\[
\begin{aligned}
  &f_k, && g_{k,1}:=k\,s_{k,k},\\
  &g_{k,y}:=-s_{k,k-1}, && g_{k,x}:=t\,g_{k,1}-a\,g_{k,y},
\end{aligned}
\]
where $f_k$ is primitive and square-free, such that, except for a finite number of $w_0 \in \C$, the
solutions of level $k$ of $\{P(x,y;w_0)=0,Q(x,y;w_0)=0\}$ are the points
\[
  \left(x=\frac{g_{k,x}}{g_{k,1}}(\tau),~ y=\frac{g_{k,y}}{g_{k,1}}(\tau)\right) \text{~for $\tau$ s.t.} ~ f_k(\tau;w_0)=0 .
\]
For each level $k$, $\{f_k,g_{k,1},g_{k,x},g_{k,y}\}$ is the
parametrisation of Corollary~\ref{coro:rur} for the system in the variables $(t,y)$. It is read directly from the triangular decomposition, without inverting $g_{k,1}$ modulo $f_k$, and all its polynomials have coefficients in $\Z[w]$, not in $\Q(w)$.

\subsection{Images at a specialisation}

The coefficients $s_{k,j}$ are minors of the Sylvester matrix of $P$ and $Q$, hence polynomials in $w$. They commute with the specialisation $w\mapsto w_0$ and the reduction modulo a prime $p$ as long as the degrees of $P$ and $Q$ in $y$ do not drop. Their degree in $w$
is bounded by 
$$B:=\deg_y(Q)\deg_w(P)+\deg_y(P)\deg_w(Q).
$$

The black box returns the factor $G_k$ of the resultant $R_y:=\operatorname{Res}_y(P,Q)$ monic in $t$ whose roots have level $k$. Since the primitive polynomial $f_k$
divides $R_y$ in $\Z[w][t]$, $\LC_t(f_k)$ divides
$\LC_t(R_y)$, and $\LC_t(R_y)\,G_k$ is the image of a polynomial of $\Z[w][t]$
of degree at most $B$ in $w$. At a pair $(p,w_0)$, we therefore record
\[
  \LC_t(R_y)(w_0)\,G_k \quad\text{and}\quad s_{k,k-1}+k\,s_{k,k}\,y \pmod p
\]
for each level $k$, both read on the output of the black box.

\subsection{Shape, separating form and bad specialisations}

The \emph{shape} of a specialisation is the list $[\deg_t (R_y),(k,\deg_t (G_k))_k]$. At a random prime $p_0$ and a random $w_0$, $a$ is the first value of $0,-1,1,-2,2,\dots$ such that every level is generic (one solution above each root) and such that the asymptotic factor of the decomposition is constant, so that no solution is hidden above a root of
$\gcd(\LC_y(P),\LC_y(Q))$. The shape at $(p_0,w_0)$ is the reference.

A degenerate specialisation can only lose solutions (to infinity or by merging) or move them to higher levels; its shape is then smaller in the lexicographic order. A pair $(p,w_0)$ whose shape is smaller than the reference is discarded; a pair whose shape is larger shows that the reference was degenerate, and replaces it. Pairs where $\deg_y(P)$ or $\deg_y(Q)$ drops are
discarded as well.

\subsection{Interpolation and multimodular lifting}

\begin{enumerate}
\item For a prime $p$ of $62$ bits, the images at $w_0=1,2,\dots$ (good pairs only) are
  interpolated in $w$ modulo $p$, coefficient by coefficient. The interpolation is exact after
  $B+1$ points; an early stop is tested on two extra points.
\item The images modulo several primes are combined by the Chinese remainder theorem. In the
  \emph{Monte Carlo} version, the lifting stops when one more prime leaves the result unchanged.
\item In the \emph{Las Vegas} version, the subresultant coefficients are interpolated on $B+1$ points and lifted with respect to the Hadamard bound of the
  Sylvester matrix. The factors $f_k$, which come from gcds modulo $p$, are then checked by exact
  divisions in $\Z[w][t]$, following the certified bivariate decomposition of the
  non-parametric case: the square-free part of $R$ is covered by the $f_k$ up to the roots of
  $\LC_y(P)$ and $\LC_y(Q)$, $f_k$ is coprime to $s_{k,k}$ and divides
  $s_{j,j}$ for $j<k$, and, for $k\ge2$, the derivatives
  $\partial^i P/\partial y^i$ and $\partial^i Q/\partial y^i$, $i<k$, vanish at
  $y=g_{k,y}/g_{k,1}$ modulo $f_k$, so that the gcd above each root of $f_k$ is a $k$-th power.
  A failed check restarts the computation.
\end{enumerate}
The parametrisation is finally made primitive in $\Z[w][t]$. All the polynomials are
handled through their images modulo primes; no Gr\"obner basis is computed.

\subsection{Description of the norm}

For $w$ generic, $\|G\|_\infty^2$ is the largest of the following \emph{candidates}: the
values $g_{k,y}/g_{k,1}$ at the real roots $\tau$ of $f_k$ with $g_{k,x}/g_{k,1}(\tau)>0$, the real
roots of $c(0,y)$ ($\omega=0$) and those of $\LC_x(c)$ ($\omega\to\infty$). These
candidates only change at the real roots of the polynomials
\[
  \LC_t(f_k),\ \operatorname{Res}_t(f_k,f_k'),\ \operatorname{Res}_t(f_k,g_{k,1}),\
  \operatorname{Res}_t(f_k,g_{k,x}),
\]
which respectively detect a solution going to infinity, two solutions meeting, a solution
leaving the parametrisation and a solution crossing $x=0$, together with the leading
coefficients and discriminants of $c(0,y)$ and $\LC_x(c)$. All of them are bivariate
resultants in $(t,w)$ or $(y,w)$. They are not factored: their primitive square-free parts are
split by pairwise gcds into pairwise coprime polynomials, whose real roots are isolated and
refined until the isolating intervals are disjoint. The parameter line is thus cut into
intervals on which the list of candidates is constant; at a rational sample point of each
interval, the specialised parametrisation gives this list, and $\|G\|_\infty^2$ is, on the
whole interval, the largest $y$-coordinate of these candidates.

This description is not cylindrical: two candidates may exchange their order inside an
interval, where two peaks of $\bar\sigma(G(i\omega))$ have the same height. Such crossings are
the double roots in $y$ of the projection $\operatorname{Res}_t(f_k,g_{k,1}y-g_{k,y})$; they are needed
to name, on each interval, the branch which realises the norm, but not to describe the norm itself.

\subsection{Stability}

The description above gives the $L_\infty$-norm; it is the $H_\infty$-norm only when $G$ is stable. Let $d(s;w)$ be the least common denominator of the entries of $G$, and write
$d=F(s^2)+s\,H(s^2)$. By the Li\'enard--Chipart criterion, $d(s;w_0)$ is Hurwitz if and only if its
coefficients and the signed principal subresultant coefficients of $(sH,F)$ ($\deg d$ odd) or
$(F,H)$ ($\deg d$ even) are nonzero and of the sign of its leading coefficient. These
polynomials of $\Z[w]$ cut the parameter line into intervals on which $G$ is either
stable or unstable, decided at one rational point of each interval. The norm is computed on the
stable intervals; at their bounds, a pole of $G$ lies on the imaginary axis and the norm is
infinite.

\subsection{Extrema of the norm}

On each stable interval, $w\mapsto\|G\|_\infty^2$ follows, piecewise, the real branches of the
plane curve $\{R_x(w,y)=0\}$, where $R_x$ is the eliminant in $y$ of $\{P,Q\}$ (eliminating $x$),
multiplied by $c(0,y)$ and $\LC_x(c)$. A local extremum occurs either at a smooth point
with a horizontal tangent, or at a corner, where two branches cross, which is a singular point
of the curve. In both cases
\[
  R_x(w,y)=0,\qquad \frac{\partial R_x}{\partial w}(w,y)=0 .
\]
This system characterises the extrema, but it is not used to compute them: the degrees of $R_x$
in $w$ and $y$ are much larger than those of the parametrisation, and grow quickly with the size
of $G$.

The extrema are read instead on a \emph{cylindrical} description of the norm. The parameter line
is also cut at the crossings of the candidates: the real roots of the discriminants in $y$ of the
projections $h_k:=\operatorname{Res}_t(f_k,g_{k,1}y-g_{k,y})$, of $c(0,y)$ and of $\LC_x(c)$, and
of their pairwise resultants in $y$. On each cell, the norm is then realised by a single branch:
a real point of level $k$, or a root of $c(0,y)$ or of $\LC_x(c)$. The stationary points of a
level are computed in the variables $(w,t)$ of the parametrisation, as the real solutions of
\[
  f_k=0,\qquad
  \frac{\partial f_k}{\partial t}\,\frac{\partial}{\partial w}\Big(\frac{g_{k,y}}{g_{k,1}}\Big)
  -\frac{\partial f_k}{\partial w}\,\frac{\partial}{\partial t}\Big(\frac{g_{k,y}}{g_{k,1}}\Big)=0
\]
(the numerator of the second equation), a bivariate system with the degrees of the
parametrisation; those of $c(0,y)$ and $\LC_x(c)$ are the solutions of $\{c=0,\partial c/\partial
w=0\}$. A stationary point is kept when it lies on the branch which realises the norm at its
value of $w$. The minimum of the norm is the smallest of its values at these points, at the
bounds of the cells and at the ends of the stable intervals; the maximum is obtained in the same
way. The candidates are exact; the values of the norm at them are evaluated numerically. The
projections $h_k$, resultants in three variables $(t,y,w)$, are only needed for this cylindrical
description.

%% file: sec_experiments.tex
The algorithm is implemented in {\sc Maple~2026} on top of the library {\tt LACE}, used through its {\sc Maple}
interface (variant linked with {\sc FLINT}) for the bivariate triangular decompositions modulo a prime,
the bivariate resultants and the real root isolation. We compare it with the algorithm of
\cite{QRY2024} restricted to one parameter, implemented with the same tools and in the same
variables $(x,y)=(\Omega,\Gamma)$: the eliminant $R_y$ in $y$ is computed from the subresultant
sequence with respect to $x$, the cells are given by $\LC_y(R_y)\,\operatorname{Res}_y(R_y,R_y')$, and
the norm is read at one sample point per cell. All timings are wall-clock times of a single run
on an Apple M2 Max with $64$\,GB of memory, {\tt LACE} using its default number of threads. All the
computations use the Las Vegas version of Section~\ref{sec:algorithm}, whose output is certified.
With respect to the Monte Carlo version, which stops the lifting much earlier, the certification
multiplies the time of the parametric RUR by about $3.5$ on the multivariable chains for $n=9$, but
increases the total time by only $17\,\%$, the computation of the cells being dominant.

Every computed value was checked against an independent frequency sweep of
$\bar\sigma(G(i \, \omega))$ at the sample points; the relative difference never exceeded
$6\cdot10^{-7}$, except on example E3 at $k_t=279$, where the sweep misses a resonance peak of
width about $10^{-3}$ at $\omega\approx20.49$: a direct evaluation at this frequency gives
$\bar\sigma^2=12.34802741$, the value computed by both methods.

\subsection{Examples from the literature}

We gathered $26$ examples with one free parameter, the other ones being fixed to rational
values: $16$ from the robust control literature (A1--D2, \cite{Doyle,Zhou,rance2018commande,QRY2024})
and $10$ physical models given by state-space representations (E1--F3), up to size $4\times4$ with
$8$ states. Among them, C3 is a two-mass/spring/damper system, E7 a quarter-car active suspension,
and F1 and F3 are chains of $3$ and $4$ masses. The systems $\{P=0,Q=0\}$ have at most $16$
solutions, and both the description of Section~\ref{sec:algorithm} and \cite{QRY2024} take less
than $0.1$\,s on every example, except F1 with \cite{QRY2024} ($0.23$\,s) and F3 ($0.36$\,s against
$1.35$\,s). The description has $140$ cells in total, against $170$ for \cite{QRY2024}; its
cylindrical variant, which also cuts at the crossings of the candidates, gives $57$ cells on A1--D2,
against $47$ and $52$. The parameter ranges are those where the systems are stable; they are
contained in the stability intervals computed by the Liénard-Chipart test of
Section~\ref{sec:algorithm}, which are often larger (for instance $\R_y$ for B2 and C2,
$(-0.4834,\infty)$ for C3).

\subsection{Order of the variables}

When $G$ is a scalar transfer function, $c$ has degree $1$ in $y$, $y=N(x)/D(x)$ is an explicit
rational function of $x$, and the elimination of $y$ is trivial, whereas \cite{QRY2024} eliminates
$x$, whose degree grows with the order of the system. These examples therefore illustrate the
effect of the order of the variables rather than the parametrisation itself.
Table~\ref{tab:siso} reports a chain of $n$ unit masses linked by unit springs and by dampers of
coefficient $w$, with a damper $1/5$ to the wall, from the force on the first mass to the
position of the last one; $P$ has degree $(2n,1,2n-2)$ in $(x,y,w)$.

\begin{table}[t]
\caption{Scalar chains of $n$ masses: cells and time (seconds) for the description of Section~\ref{sec:algorithm} (RUR), for its
cylindrical variant and for \cite{QRY2024}.}\label{tab:siso}
\centering\small
\begin{tabular}{rrrrrrr}
\hline
$n$ & \multicolumn{3}{c}{cells} & \multicolumn{3}{c}{time} \\
 & RUR & RUR cyl. & \cite{QRY2024} & RUR & RUR cyl. & \cite{QRY2024} \\
\hline
2 & 3 & 7 & 7 & 0.01 & 0.02 & 0.04 \\
3 & 7 & 17 & 16 & 0.04 & 0.06 & 0.19 \\
4 & 10 & 29 & 28 & 0.05 & 0.21 & 0.74 \\
5 & 13 & 43 & 42 & 0.06 & 0.60 & 3.75 \\
6 & 16 & 60 & 59 & 0.09 & 1.66 & 13.26 \\
7 & 19 & 83 & 82 & 0.11 & 4.34 & 39.80 \\
8 & 22 & 113 & 112 & 0.15 & 12.63 & 121.19 \\
\hline
\end{tabular}
\end{table}

\subsection{Multivariable chains}

The same chain of $n$ masses with forces applied to, and positions measured on, all the masses gives $$G=(s^2 \, I+s \, B+K)^{-1},$$ an $n\times n$ symmetric matrix; $P$ has degree $(2n,n,2n-2)$ and
$\{P=0,Q=0\}$ has $n^2$ solutions. Table~\ref{tab:mimo} details the cost of each step. The
parametric RUR itself remains cheap; most of the time is spent in the resultants defining the
cells, whose degree in $w$ reaches $3912$ for $n=9$. The algorithm of \cite{QRY2024} took $3.4$,
$24.9$ and $194.6$\,s for $n=4,5,6$ (measured in the same session as the plots below), giving
$88$, $176$ and $320$ cells.

\begin{table}[t]
\caption{Multivariable chains, $G$ of size $n\times n$: number of solutions, degree in $w$ of the
parametric RUR, and time (seconds) of its computation, of the critical polynomials in $w$ (whose
real roots bound the cells), and of the sample points; number of cells.}\label{tab:mimo}
\centering\small
\begin{tabular}{rrrrrrrr}
\hline
$n$ & sol. & $\deg_w f$ & RUR & critical & samples & total & cells \\
\hline
4 & 16 & 30 & 0.3 & 0.3 & 0.1 & 0.6 & 54 \\
5 & 25 & 48 & 1.2 & 1.6 & 0.4 & 3.2 & 87 \\
6 & 36 & 70 & 4.8 & 8.5 & 1.5 & 14.8 & 145 \\
7 & 49 & 96 & 18.2 & 35.2 & 6.1 & 59.5 & 224 \\
8 & 64 & 126 & 55.3 & 126.8 & 23.6 & 205.6 & 326 \\
9 & 81 & 160 & 145.3 & 467.7 & 69.8 & 682.8 & 450 \\
\hline
\end{tabular}
\end{table}

\subsection{Plotting and extrema}

\begin{figure*}[t]
\centering\includegraphics[width=\textwidth]{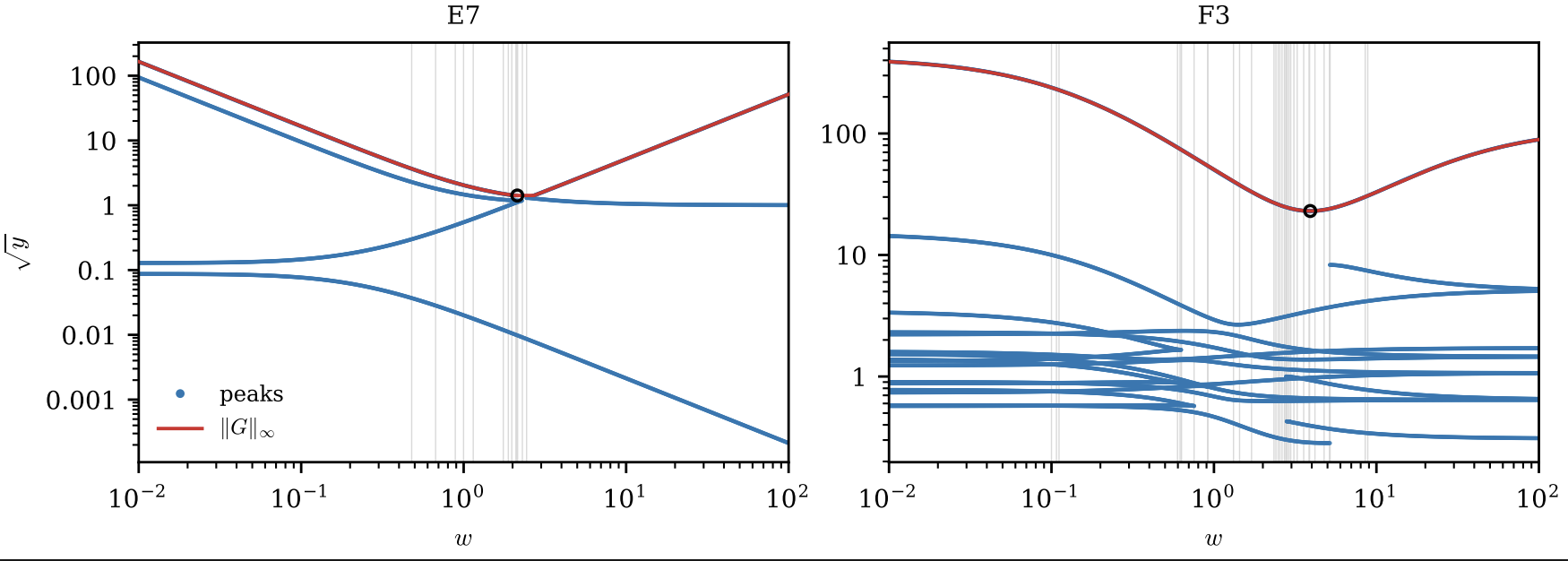}
\caption{Critical curves and norm on E7 and F3, from the description of Section~\ref{sec:algorithm}
at $800$ points: the peaks $\sqrt y$ of the real solutions with $x>0$ (blue), their maximum
$\|G\|_\infty$ (red), the bounds of the cells (vertical lines) and the minimum of
Table~\ref{tab:extrema} (circle). On E7, the norm passes from one branch to another, and is
equal to its minimum $\sqrt2$ on the whole interval between them, about $[2.10,2.68]$.}\label{fig:curves}
\end{figure*}

For plotting $w\mapsto\|G\|_\infty$, the parametrisation is specialised
at each point of the grid. Fig.~\ref{fig:curves} shows the critical curves and their maximum. On F3 and $400$ points, this takes $0.91$\,s after a preparation of
$0.46$\,s, against
$1.49$\,s for solving the non-parametric problem at each point, and $1.42+1.49$\,s with
\cite{QRY2024}. For the multivariable chains with $200$ points, the specialisation is about
$25\,\%$ faster than the direct solution ($2.18$ against $2.80$\,s for $n=6$), but its
preparation is only paid back after a few thousand points: the interest of the description is
the exact location of the cells rather than the speed of the plot.

Table~\ref{tab:extrema} reports the minimum over $w$ of the norm, computed from the cylindrical
description of Section~\ref{sec:algorithm} (RUR cyl.) and from the cylindrical description of
\cite{QRY2024}, whose branches are those of its eliminant $R$ and whose stationary points are the
real solutions of $\{R=0,\partial R/\partial w=0\}$. Both give the same values. The minimum is
the minimum of one resonance peak, except for E7, where the norm is equal to $\sqrt2$ on a whole
interval of $b_s$, and for the chain with $k=2$, $n=6$, where it is reached at a bound of a cell; the parametrisation also gives the frequency where it is reached. On the examples \blue{in} the 
literature, \cite{QRY2024} is faster on F1 and F3, where the projections $h_k$ in three variables dominate. On the chains, the description of Section~\ref{sec:algorithm} is faster, and the gap grows with $n$: \cite{QRY2024} did not finish in $30$ minutes for the scalar chain with $n=7$, nor in $15$ minutes for the multivariable chains with $n=5$.

\subsection{Link to the code}
The implementations used for these experiments can be found on the "Experimental" section of the PACE project website at  \url{https://pace.gitlabpages.inria.fr/tutorials/}

\begin{table}[t]
\caption{Minimum of $\|G\|_\infty^2$ over the parameter: value, parameter and frequency where it is reached, time (seconds) from the cylindrical description of Section~\ref{sec:algorithm} (RUR
cyl.) and from that of \cite{QRY2024}. Chains: scalar ($k=1$) or $k\times k$, with $n$ masses.
$^\dagger$: measured while two other computations were running.}\label{tab:extrema}
\centering\footnotesize\setlength{\tabcolsep}{3pt}
\begin{tabular}{lrrrrr}
\hline
 & $\min\|G\|_\infty^2$ & $w^\ast$ & $\omega^{\ast2}$ & RUR cyl. & \cite{QRY2024} \\
\hline
C3 & $12.92274436$ & $6.290606$ & $0.8482076$ & $0.18$ & $0.16$ \\
E7 & $2$ & -- & -- & $0.17$ & $0.32$ \\
F1 & $95.56211538$ & $3.305428$ & $0.2901666$ & $1.45$ & $0.78$ \\
F3 & $529.7247039$ & $3.913804$ & $0.1593434$ & $21.6$ & $5.34$ \\
\hline
$k=1$, $n=4$ & $7.584151832$ & $3.802189$ & $0.1698369$ & $0.32$ & $2.80$ \\
$n=5$ & $5.693192393$ & $4.836474$ & $0.1224024$ & $0.81$ & $37.9$ \\
$n=6$ & $4.584484756$ & $5.955679$ & $0.0928166$ & $2.12^\dagger$ & $426^\dagger$ \\
$n=7$ & $3.871981609$ & $7.153454$ & $0.0729841$ & $6.00^\dagger$ & $>1800$ \\
$n=8$ & $3.381934435$ & $8.425143$ & $0.0589823$ & $29.3^\dagger$ & -- \\
\hline
$k=2$, $n=2$ & $82.97580527$ & $2.270489$ & $0.4394113$ & $0.15$ & $0.26$ \\
$n=3$ & $56.04551617$ & $3.419953$ & $0.2644210$ & $1.17$ & $2.56$ \\
$n=4$ & $43.14098715$ & $4.849875$ & $0.1839832$ & $7.37$ & $56.9$ \\
$n=5$ & $36.34200585$ & $6.514696$ & $0.1394068$ & $35.3^\dagger$ & $>900$ \\
$n=6$ & $38.34807683$ & $6.413239$ & -- & $332^\dagger$ & $>900$ \\
\hline
$k=3$, $n=3$ & $129.0126117$ & $3.464448$ & $0.2658412$ & $1.58$ & $1.57$ \\
$n=4$ & $107.1194709$ & $5.121686$ & $0.1885347$ & $28.0$ & $95.0$ \\
$n=5$ & $82.82399684$ & $6.519991$ & $0.1403110$ & $349^\dagger$ & $>900$ \\
\hline
\end{tabular}
\end{table}